The Uranus Flagship: Investigating New Paradigms for Outer Solar System Exploration
Workshop Summary Report

**Conveners:** Amy Simon, Louise Prockter, Ian Cohen, Kathleen Mandt, Lynnae Quick

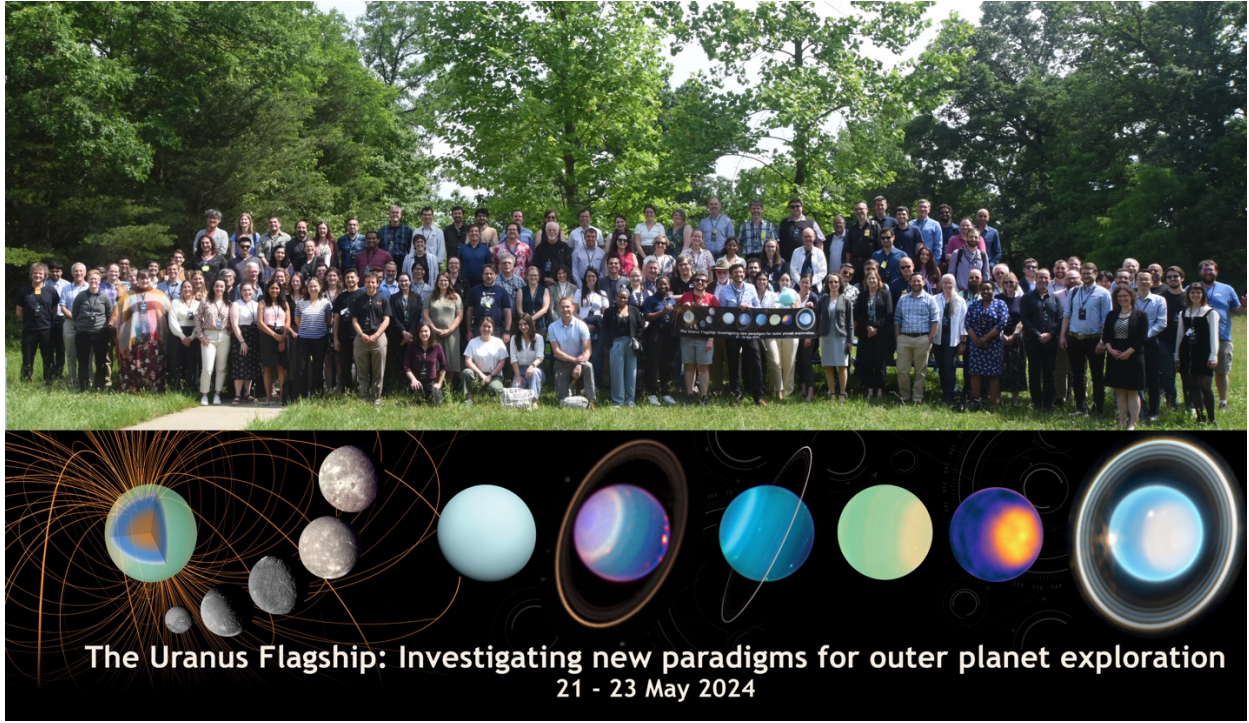

## Goals

This Workshop took place 21 to 23 May 2024 at NASA Goddard Space Flight Center, about a year after the successful July 2023 Pasadena workshop on investigations and instruments for cross-discipline science on the Uranus Flagship.  Co-led by GSFC and APL conveners, we had a broad, international, Science Organizing Committee, and a largely early career Local Organizing Committee from APL and GSFC. From the 2023 meeting, it was apparent that the community was wildly enthusiastic about starting a mission, but lacked focus on what was possible or where to begin.  Thus, the purpose of our workshop was to discuss practical aspects of the next planetary flagship and how we can employ new paradigms to better enable robust outer planet exploration.  To enable this goal, we introduced the community to the best practices and lessons learned from previous missions and NASA-commissioned studies, and discussed the challenges involved with a mission so far from the Earth/Sun.  The underlying workshop purpose was to steward the community towards a more practical mission design approach that will enable the development of this mission, as well as future missions, on a shorter cadence by setting expectations and having difficult discussions early in development. Because of the time scales involved in this mission, special effort was made towards early career inclusion and participation.

# Table of Contents

## Workshop Format and Topics

The Workshop spanned 3 days, with sessions consisting of plenary talks, followed by Q&A (online attendees could add questions via Slido). Most speakers were invited to cover specific topics, but some were chosen from submitted abstracts. Day 1 topics were focused on Lessons Learned and Best Practices, with sessions on mission costing, enabling interdisciplinary and cross-divisional science, and early career training opportunities (see session list below). The morning of Day 2 introduced the practical challenges of exploring at 20AU starting with mission design with a focus on mission architecture, instrument accommodation, probe design, and trajectories. The remaining talks were from speakers chosen from abstract submission, but with a post-selection mandate to look at a single Decadal study mission objective and to list all the possible (instrument agnostic) ways to answer the objective, prioritized, and with a list of how they drive mission design. This was followed by a broad poster session. Day 3 of the workshop split attendees to activities of their choosing. At GSFC, participants listened to lectures on mission development lifecycles and science traceability, followed by a cost-constrained mission design activity, with optional tours of GSFC facilities in the afternoon. Other participants traveled to Washington, DC to engage in science advocacy activities led by APL.

## Meeting Statistics

More than 275 people registered for the Workshop, with a full ~2/3 intending to attend in person. Owing to badging restrictions and foreign national escort requirements, in person registration was not allowed, but people from other NASA Centers could attend if their badge allowed GSFC physical access. Even with some people unable to attend, we consistently saw about 165 people in the meeting rooms, with a varying number participating in the live stream through the sessions. We had participants from many other countries including France, Italy, Ireland, Japan, Germany, and England, some representing other space agencies, including DLR, JAXA, and ASI. As part of registration, we also asked how many would be interesting in hearing about early career events, and ~120 responded, including 80 of the in-person attendees.

Although our call for abstracts was very focused on near-term technologies, interdisciplinary science, and science mission drivers, we still received more than 80 abstracts (not all fully responsive to the call). Of these, some were merged and elevated into talks, resulting in about 65 posters at the meeting for the Day 2 afternoon poster session.

## Early Career Participation

To encourage early career participation, we first applied to the TWSC program for early career travel support (where early career was defined as <10 years post PhD). To apply, each applicant sent in a proposal package consisting of a statement of what they hoped to achieve at the workshop, a CV, a letter of endorsement from a supervisor or colleague, and an estimate of travel expenses. We had 19 applicants and were able to select 14 for awards. Award criteria included giving preference to those from traditionally unrepresented institutions, as well as recipients who did not receive an award in 2023. Our awardees were: C. Gentgen (MIT), L. Guan (U. Houston), A. Conly (NMSU), V. Afigbo (U. Idaho), E. Dahl (JPL), A. Denton (U. Arizona), K. Davis (FIT), A. Walker (Howard), A. Bryant (U. Chicago), G. Miceli (UC. Boulder), P. Johnson (UT Austin), H. George (UC Boulder), Y. Phal (CO School of Mines), K. Miller (SwRI San Antonio), D. Qasim (SwRI San Antonio, withdrew). Many of the recipients are students, including undergraduates, and we had a high proportion of underrepresented minorities.

In addition to travel support, we also offered several early career networking events. These included an optional peer Happy Hour organized by the Local Organizing Committee, and a catered networking breakfast sponsored by APL. For the breakfast, we had 30 early career attendees who met with more senior community members from APL, JPL, and GSFC, using an ice breaker activity designed to facilitate discussion.

## Poll and Survey Results

Slido polls were conducted during the workshop, including demographics (~72 respondents) and fun topics. An optional post-workshop survey was also conducted to receive feedback for future conveners. All 30+ respondents expressed satisfaction with the meeting.

**Career stage:**
- Undergraduate student - 1%
- Graduate student - 10%
- Postdoc - 15%
- Early-career Professional (<10 yrs from terminal degree) - 35%
- Professional (10-19 yrs from terminal degree) - 22%
- Professional (>20 yrs from terminal degree) - 17%

**Career Sector:**
- Academia - 43%
- For profit - 0%
- Government - 24%
- Nonprofit - 14%
- Not sure - 3%
- I'm a student - 10%
- Other - 7%

**Funded to work on Uranus:** Yes - 13%, No - 88%

**Non-binding, unofficial, name poll:**
- A mythological figure (like Juno) - 40%
- A Shakespearean character - 22%
- An historical astronomer (like Cassini) - 18%
- A more recent scientist (like Roman) - 4%
- An acronym (like OSIRIS-REx) - 16%

**Write-In Name Poll (Top 3):**
    Tempest: violent storm
    Caleus: Roman name for Uranus, could also be an acronym
    Hypatia: 4[th] century female astronomer, philosopher, and mathemetician

**Post-Workshop Survey (subset):**

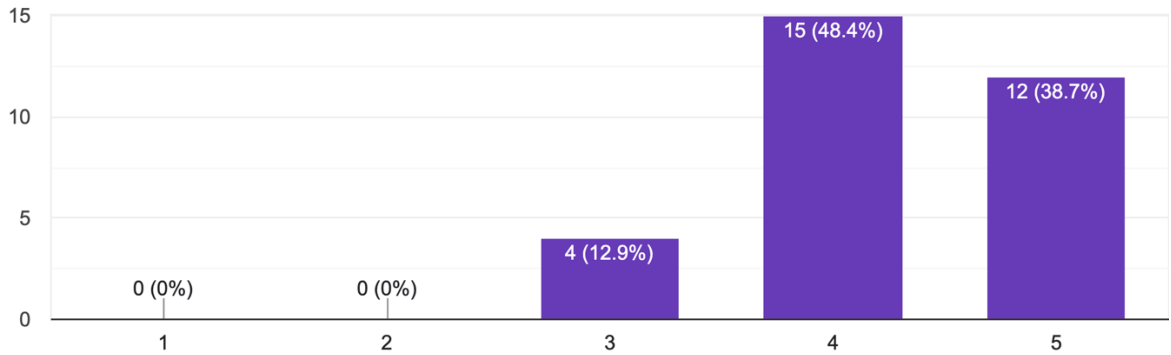

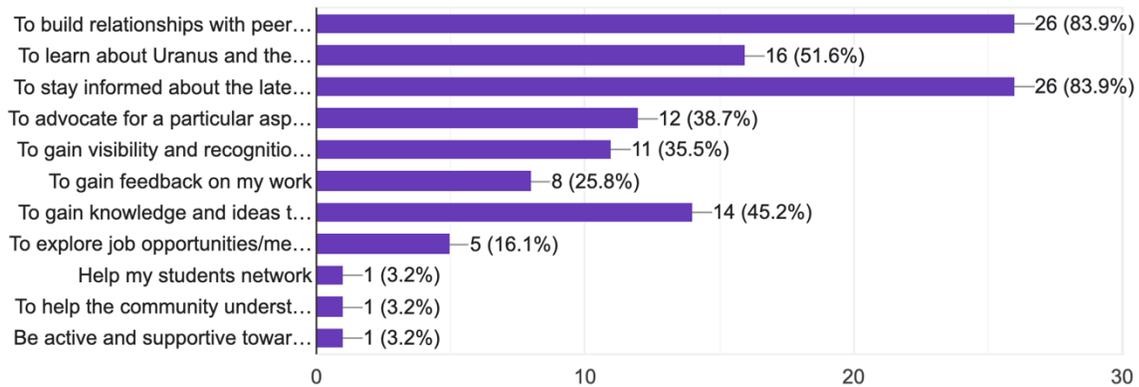

## Session Summaries and Key Points

### Session 1: Mission Costs

**Speakers**: R. Persinger, V. Hamilton, O. Figueroa, J. Crooke, A. Driesman

**Summary**: This session discussed mission costing, with an aim of explaining why mission costs are much higher than the community anticipates, but also lessons learned to avoid, or limit, cost growth. Several speakers noted that independent technical, management, risk, and costs reviews are essential; mission advocate costs are often underestimated, especially during rapid Decadal studies. It is also recognized that a Flagship/Strategic mission addresses the highest priority Decadal science, enables broader and deeper investigations than a competed mission, maintains US leadership in space and science, has a high degree of external visibility, and offers greater opportunities for international participation, cooperation, and collaboration (Hamilton).

The UOP Decadal concept study was assessed as low-medium risk, in part because it limited the new technology required, and because it attempted to constrain the science requirements (Persinger). The Aerospace-identified biggest growth risks are payload selection, including instrument development risk and schedule, RTG availability, orbiter/probe mass growth, total $\Delta V$ and propellant (dependent on interplanetary trajectories and tour), and science tour definition and duration. All of these are expected to be refined as the mission progresses.

**Other Key Points**:
- The cost of large missions is understandably of considerable interest vs. smaller missions, and cost growth can have a significant impact across the Division.
- Must provide sufficient funds up front to retire risks definitively prior to confirmation (KDP-C).
    - Don't undersell the early mission formulation/development needs.
- Allow some flexibility as design matures.
    - Poor decisions may drive the design too hard in the wrong direction.
    - Allow for science discoveries and synergies in operations.
- Early requirements definition will aid instrument choices and accommodation.
- Limit unnecessary new technologies.
    - Need adequate investment in Next Gen RTGs.
    - Instruments may need development.
- Consider team building and inclusion from the start.
    - Can't plan for the unknown but can build a resilient team.
    - Have strong succession and retention planning across all elements.

## Session 2: Enabling Interdisciplinary & Cross-Divisional Science

Speakers: S. Brooks, M. Dougherty, G. Portyankina, I. Cohen, L. Mayorga

**Summary**: This session discussed aspects of what is needed to enable interdisciplinary and cross-divisional science. Cassini is a prime example of a planetary science mission that incorporated interdisciplinary science into operations planning. Implementation was led by science planning that incorporated input from multiple disciplines (e.g. Rings, Titan, Magnetospheres, etc.) to ensure that the full range of science goals were achieved (Brooks). Cassini team relationship development during the long cruise included weekly operations planning telecons that established an environment for collaborative and interdisciplinary science at Saturn (Dougherty).

When planning observations for the ESA JUICE mission, prioritization of Regions of Interest (ROI) provided a mechanism for exploiting the full capabilities of the JUICE instruments in achieving science goals requiring multiple instrument measurements (Portyankina). Finally, high priority science goals relevant to Astrophysics and Heliophysics Divisions could be achieved with this mission if the Divisions coordinate during mission design. Studies of the solar wind (Cohen) and analog observations of Uranus as an exoplanet (Mayorga) could be achieved during transit with the appropriate instrument suite. In fact, exoplanet scientists have a well-defined "wish list" after arrival at Uranus that would have a major impact on the field (Mayorga).

**Other Key Points**:
- Large missions tend to have distributed operations with teams across the US and internationally. Planning tools and ground system architecture should be designed with this in mind, and spacecraft commanding should not be highly interactive.
- A spacecraft without a scanning platform requires a much larger investment in operations planning. Europa Clipper used this lesson learned to place instruments so that they were optimized for nadir-pointed flybys.
- Don't plan science operations so definitively that room is not allowed for unexpected science discoveries.
- Cross-instrument synergies in planetary missions have traditionally been achieved after data acquisition.
    - This should be incorporated into mission design and operations planning.
    - Ensure that cross-instrument interference does not prevent simultaneous operations.
- Team building
    - Ensure a variety of skill sets and knowledge.
    - Include succession planning from the start and training of early career researchers.
    - Build relationships across instrument teams and discipline working groups.

## Session 3: Early Career Opportunities and Training Programs
Speakers: D.J. Smith, S. Klug-Boonstra, B. Fernando

**Summary:** To ensure a future diverse work force, we need trained scientists and engineers who have early opportunities to shadow missions and have mentors.  In addition, we need to reach out to disadvantaged groups and "meet them where they are," rather than assuming everyone has had a level playing field and opportunities until this point.  It is also important that skills that are learned are more broadly applicable than just to planetary scientists/engineers, as it is critical to have a technically educated and trained workforce across all employment sectors.  In this session we heard examples and program details from L'Space, Here2Observe, and the Insightseers programs.

L'Space has been wildly successful at reaching non-traditional audiences.  The program is free, does not require being at a specific participating institution, and has a lower GPA requirement than traditional programs.  This program is designed more about training a future work force, starting mostly at the undergraduate level.  Here2Observe is a direct shadow opportunity where university faculty apply to host a cohort paired with a specific NASA mission, again at the undergraduate level.  Insightseers is an early career participation program, targeted to the graduate student/postdoc level, where applicants are paired with a mentor and attend Science Team Meetings.

**Other Key Points**:
- To reach non-traditional students, you have to reset expectations and keep the time commitment reasonable.  "Meet them where they are."
    - Not all of these students will have an interest in planetary science or NASA missions, so skills need to be broadly applicable.
    - Self-guided, self-paced, modules are helpful.
    - It is still crucial to have opportunities for internships.
- At the professional early career level in-person participation is preferred in programs such as Insightseers Science Team Meetings, which is critical for networking.
- They may still be a gap in retention of early (and mid) career community members.
    - Planetary is generally a soft money field, with no guarantee of future mission opportunities.
    - Need strong training and succession planning across all aspects of a mission.
    - Need to include normalizing "emeritus" membership in missions to encourage succession occurs.
    - Need to encourage flexibility in early career scientists, as well, so they can pursue the opportunities that are available, rather than waiting for a narrow field to emerge as a priority for funding.

## Session 4: Mission Design and Challenges for Uranus

**Speakers:** R. Anderson, S. Feldman, M. Amato, D. Ellison, J. Englander, D. Landau, K. Hughes

**Summary:** Exploring at 20AU is not trivial. Power options are limited, and resources, including deliverable mass, returnable data volume, and available power will be constrained. Since the Decadal survey, some parts of the mission concept have changed. Based on current possible launch dates, the Jupiter gravity assist is unlikely to be available, affecting flight time and deliverable mass. The trade space now includes inner solar system gravity assists, solar electric propulsion, or newer launch vehicles, as they become available.

Additionally, the availability of 3 NextGen Mod 1 RTGs seems unlikely, but a reasonable mission can close on 2 RTGs, with some consequences on mission flexibility (instruments cannot be powered on 100% of the time). Orbiter instruments will need to be chosen with this in mind, as well as pointing directions for competing objectives. On probe design, accommodation and packing depends strongly on the choice of payload. Studies are underway to reduce the requirements (deceleration loads) on the instruments.

Other Key Points:
- The mission closes with existing technology, even with later launch dates.
    - Aerocapture is not required, and would require deep study of thermal accommodation, aeroshell integration, and packaging of a large spacecraft, with a probe and RTGs.
    - NextGen mod 1 RTGs and the availability of plutonium remain a concern.
    - All data can be returned using Ka-band communication.
- Trajectories with flight times of 10-12 years are possible.
    - Launch dates and cruise options need to be defined.
    - The ring environment remains a challenge – likely orbit insertion will occur farther from the planet than in the Decadal study.
    - Close periapses passes of the planet are also a challenge unless the spacecraft moves away from the equatorial plane.
    - Orbital tours can be optimized for multiple science objectives, once payloads are chosen.
- Early instrument definition will help to prevent accommodation issues.
    - Important to have required instrument engineering units (rather than descopable, fully assembled, flight spares which are expensive).
    - Operations will have to include multiple objectives, and flexibility for new discoveries.

## Sessions 5 and 6: Science Optimization and Mission Drivers

**Introduction:** Speakers for Sessions 5 and 6 were chosen from abstract submissions and to ensure coverage across the UOP Decadal study objectives; some abstracts were merged where two submissions better covered the topic. While not all objectives could be discussed, all major disciplines were included across the breadth of the 10 selected talks. To ensure that the topics were covered evenly and met the Workshop goals, speakers were given an optional template and instructed to cover the following for their objectives:

For each objective:
- List of measurements in the Decadal STM.
- Any additional methods/measurements to add.
- Keep this instrument agnostic at this point, to the extent possible.

For each measurement:
- Requirements
    - Instrument/science needs (type(s) of instruments, spatial res, probe depth, etc).
    - Mission drivers (number of flybys, distances, frequency of observation, etc).
    - Focus on the *minimum* needed to meet the objective.
- Pros and Cons of this approach
    - Does it fulfill objective in part or in full?
    - Does it uniquely address part of the objective?
    - Does it overly drive the mission architecture?

What measurements best meet objective:
- Ranked list of measurements/techniques.
- Can have several must-do's to meet the objective, but everything can't be top priority.
- Can include a table showing how they address objective together (i.e., fever or "consumer reports"-type chart).

Summary of corresponding mission drivers:

Take Away Messages:
- Best measurements to meet objectives and what they require.
- If desired: what bonus you could get from other measurements above and beyond the minimum requirements.

The next pages summarize the findings presented for each of the covered UOP Objectives.

## M1. *What dynamo process produces Uranus's complex magnetic field?*

**Speakers:** K. Soderlund et al.

**Summary**:
A particularly important question called out by the recent Planetary Decadal Survey is "What is responsible for the differences between the magnetospheres of the gas giants and ice giants?" One step in answering this is understanding the dynamo process in Uranus. Although we know that planetary magnetic fields are generated by convectively-driven dynamos in electrically conducting fluid regions, we do not understand what dynamo process produces the magnetic field of Uranus. To answer this question we need to determine what parts of Uranus are convective, the planet's bulk composition and depth dependence of composition, if distinct compositional boundaries exist, and if zonal winds interact with the dynamo. This requires measuring the internal magnetic field structure, the gravity field and ring/planetary oscillations and shape, and global energy balance.

Required Measurements:
- Spatially resolved magnetic field intensity and orientation
- Internal magnetic field variations over time
- Zonal gravity field
- Tidal Love number $k_{22}$

Lower priority:
- UV emission from auroral and satellite footprints
- IR emission from auroral and satellite footprints
- Pole orientation
- Longitudinal and temporal variations of the gravity field

Mission Drivers:
- Operational duration of at least ~4 years
- Require at least 8 orbit pericenters. Outside of the rings will achieve spherical harmonic (SH) degree and order 4, inside the rings with moderate inclination and uniform longitudinal coverage achieves SH 7-8. Ideal orbit described below achieves SH 10-12.
- Require at least 4 orbit pericenters outside of the rings to achieve up to J4 outside of the rings and J6 inside the rings with moderate inclination. Ideal orbit described below achieves gravitational moments up to at least J8 and much smaller uncertainties on J2 to J6.

Other Desires (not requirements)
- At least 8 orbits with periapses inside the rings with high inclination and uniform longitudinal coverage.

*M2. What are the plasma sources and dynamics of Uranus' magnetosphere and how do they interact with the solar wind?*

**Speakers:** R. Ebert et al.

**Summary**:

The second part of answering the question "What is responsible for the differences between the magnetospheres of the gas giants and ice giants?" is to determine the plasma sources and dynamics of ice giant magnetospheres and to understand how they interact with the solar wind. We need to determine the dominant processes governing the magnetosphere; constrain the structure, dynamics, and temporal evolution; investigate the evolving composition of the magnetosphere and planet's ionosphere; and determine the variability and thermal structure of the thermosphere and ionosphere.

Required Measurements:
- Magnetic field magnitude and direction of the upstream solar wind and near the magnetopause.
- Ion and electron measurements. A trade study should be conducted to determine ideal energy ranges and instrument requirements.

Lower priority:
- Some optimized combination of:
  - Thermal and suprathermal ion moments for energies 0.2-10 keV.
  - Electron moments for energies 0.01-10s of keV with pitch angle coverage from 0-180 degrees.
  - Energetic particles moments at energies 10s-100s keV for electrons and ions.
  - Plasma wave electric and magnetic fields Hz-MHz range
- Auroral imaging in the UV and IR

Mission Drivers:
- Operation during approach to measure upstream solar wind properties starting at minimum ~0.5 AU from Uranus with continuous observations covering several solar rotation periods. Thermal plasma instrument with solar wind in field-of-view.
- Orbit apoapsis beyond dayside magnetopause location (~30 $R_U$) with magnetopause crossings at different local times and latitudes.
- Sufficient magnetic cleanliness to make measurements.

Other Desires (not requirements)
- Auroral imaging pointing toward the auroral region when the spacecraft is near apoapsis.

## R1. *What processes sculpted the ring-moon system into their current configuration? (+R2)*

**Speakers:** M. Hedman et al.

**Summary**:

The rings and inner small moons are a complex dynamic system that requires multiple measurements to properly investigate. At this point, we do not know how the narrow rings are confined or how chaotic is the satellite system, topics we understand much better for Saturn, after Cassini. Additionally, an important question is how the ring-moon system is evolving over time, in particular, tidal forces should cause the moons to migrate inwards towards the planet, allowing material to cycle between rings and moons. High-resolution data are needed to understand the current state and evolution of the rings. Compositional data are also needed to constrain the longer-term history and origins of the rings and moons.

Required Measurements:
- Imaging of the dense rings, <100-m resolution
- Stellar and radio occultations
- Remote composition measurements
- Imaging and composition of the inner small moons, ~200-m resolution
- Temporal and longitudinal coverage of rings at a range of phase and opening angles

Lower priority:
- Remote measurements of material flux into and out of rings
- Regular surveys of the ring-moon system, ~10-km resolution

Mission Drivers:
- Orbit geometry to allow close observations of rings and small inner moons
- Occultation geometries
- Multiple observations of rings and moons over a time period of at least a year
- Range of viewing geometries (ring longitude, phase, and opening angles)

Other Desires (not requirements)
- Arrival date before equinox to see vertical structures generated by embedded objects, external perturbations
- In situ sampling of rings/material flux, if it can be done safely

*R2. What are the compositions, origins, and history of the Uranian rings and inner small moons?*

**Speakers:** A. Verbiscer, T. Denk et al.

**Summary**: Irregular moons were not included in the Decadal STM, but should be added as a component to improving understanding of the rings and inner small moons. Irregular moons are progenitor objects formed far away from the planet and captured into orbit by an unknown process. They have been exposed to heavy collisional environments and are observed in orbital "families." Close encounters of Saturn's irregular moons with Cassini provided details information about surface geology and showed unique composition compared to regular moons.

Required Measurements:
- Object tracking using a small telescope of at least one irregular moon over several hours when other observations are not being made.
- Flyby of irregular moon with imaging and spectroscopy, which is likely to be easiest during approach or first orbits.

Lower priority:
- Flybys of more than one irregular moon.

Mission Drivers:
- See required measurements.

*A1. How does atmospheric circulation function from interior to thermosphere?*

**Speaker**: E. Dahl, K. Sayanagi, et al.

**Summary**:
Uranus's atmosphere is full of mysteries: a relatively sluggish atmosphere relative to Neptune, despite higher levels of insolation, an apparent lack of internal heat release might be caused by molecular/thermal boundary in deep atmosphere, and an anomalously warm thermosphere (cooling down) and cool stratosphere. Deriving a 3D picture of atmospheric circulation in Uranus's atmosphere will help us understand the thermal evolution/energy balance of the planet, how gas/ice giant atmospheres respond to extreme seasonal forcing, meteorology, and storms (on Uranus and elsewhere), and the degree to which remote observations are representative of global/deeper composition.

Required Measurements:
- Cloud top winds
- Disequilibrium and condensable composition ($CH_4$, $H_2S$, $NH_3$, $H_3^+$, $H_2$, $C_2H_2$, $C_2H_6$, ortho-para fraction)
- In situ winds and composition
- In situ temperature-pressure profiles
- Radio Occultations (for upper atmosphere)

Lower priority:
- Deep winds (via gravity)
- Stellar Occultations
- In situ cloud measurements

Mission Drivers:
- Probe for in situ measurements, 5 bars required, 10 bars desired
- Dayside observations and observational cadence (time separated imaging for winds)
- Occultation geometries
- Highly elliptical polar passes (for gravity/deep winds) with periapse < 1.1 $R_U$

Other Desires (not requirements)
- Earth-based support
- Multiple probes
- Arrival date before equinox

*A3/I1. When, where, and how did Uranus form, and how did it evolve?*

**Speakers:** K. Mandt et al.

**Summary**: The first three questions posed decadal survey focus on improving our understanding of the formation and early evolution of the solar system. These questions call out a need to know how the composition of solid materials and gas in the protosolar nebula varied with distance from the Sun and with time, how the giant planets formed, and how they migrated after formation. Determining this has implications for our understanding of the architecture of our solar system which is the number and types of planets we have, where these planets orbit, and the distribution of small body populations. By extension, answering this question will help us to understand how Earth managed to end up in a location where liquid water could exist on the surface and how the water we have was delivered to Earth.

Required Measurements:
- Noble gas abundances relative to hydrogen in the atmosphere of Uranus.
- Noble gas isotope ratios for He and Xe and the D/H isotope ratio
- Broad wavelength Bond Albedo and thermal emission for thermal balance
- Global cloud top composition
- Ground-based microwave observations 2.6-33.5 GHz for deep composition

Lower priority:
- Noble gas isotope ratios for Ne, Ar, and Kr.
- Elemental abundances of C, N, O, and S.
- Isotope ratios of C, N, O, and S.
- Orbital microwave measurements 0.6-1.2 GHz for oxygen deep abundance.

Mission Drivers:
- An atmospheric probe with a mass spectrometer is required to measure the noble gases.

Other Desires (not requirements)
- Coordination with ground-based observations in the microwave.

*I2. What is the bulk composition and its depth dependence? & I3. Does Uranus have discrete layers or a fuzzy core?*

**Speakers**: S. Markham, M. Parisi et al.

**Summary**:

The interior structure of an ice giant is not very well known nor constrained. The deepest interior helps to drive the unique magnetic field, while the deep structure affects the convective and radiative balance in the planet. Deep circulation, and atmospheric profiles, also affect the visible weather layer. The gravity field also affects the stability of the rings and satellites. Thus, the interior structure objectives have significant overlap and synergies with atmospheric, magnetosphere, rings, and satellite science objectives, and have some complementary measurements.

Required Measurements:
- Gravity field at least to $J_8$, with improved uncertainties on $J_2 - J_6$.
- High resolution ring imaging for oscillations
- Gravitational seismology from radio telemetry

Lower priority:
- Surface velocimetry/Doppler seismology

Mission Drivers:
- Orbits with low periapse < 1.1 $R_U$, high inclination, longitude of ascending node oriented along the Uranus-Earth line-of-sight vector
- If Doppler seismology is included, large data volume requirements

Other Desires (not requirements)
- In addition to low periapse, orbits with (in priority order):
  - pericenter near the equator
  - low eccentricity
- Additional orbit geometries

*S1. What are the internal structures, rock-to-ice ratios of the large Uranian moons, and which possess internal heat sources or possible oceans?*

**Speakers:** V. Filice et al.

**Summary**:
The Uranian system possesses the last unexplored set of regular moons. The rapid Voyager flybys showed unique geologic features suggestive of interesting interior structure. Thus, we desire to know more about their internal structures and rock-to-ice ratios. There are also hints that some of these moons may be, or have been, ocean worlds. Another major objective is to determine which possess substantial internal heat sources or possible oceans today.

Required Measurements:
- Doppler tracking for gravity measurements on equatorial flyby
    - Static gravity, $k_2$ Tidal Love number
- Surface feature imaging throughout a flyby
    - Libration, obliquity

Lower priority:
- Polar flybys to improve $J_2$ on all 5 major moons
- Repeated passes to sample each satellite's mean anomaly

Mission Drivers:
- Orbits with multiple gravity passes at different orbital longitudes, low altitude
    - Must have direct to earth communication during the pass
    - Ka-band preferred
- May drive instrument orientation and power profile (to have both gravity and other remote sensing on the same passes)

*S3. What geological history and processes do the surfaces record and how can they inform outer solar system impactor populations?*

**Speakers**: C. Beddingfield, M. Kinczyk et al.

**Summary**:

We learned a great deal from Voyager 2 images of the major Uranian moons, Titania, Ariel, Miranda, Umbriel and Oberon.  Nearly all of these are past or present candidate ocean worlds as we now know them to exist at Jupiter and Saturn.  The moons experienced likely multiple past orbital resonances and repeated periods of geologic activity.  However, the drivers, timing, and inter-relationships of events remain uncertain, along with current evolution and surface processing.  Detailed imaging and composition measurements are needed to better understand these moons.

Required Measurements:
- Surface imaging
    - To determine impactor fluxes and impact crater size frequency distributions
    - To constrain drivers and ages of past geologic activity
    - 100 m/pixel global, higher resolution and stereo over geologic features
- In situ particle measurements to search for geologic evidence of past and/or present plumes
- induced magnetic field measurements to investigate possible subsurface oceans
- Surface composition measurements of salts and other endogenic species (0.1 to 1 km/pixel)

Lower priority:
- High phase imaging for plumes

Mission Drivers:
- Satellite surface coverage, especially of northern hemispheres
- Viewing angle coverage for stereo
- Flyby distance and speed will depend on chosen payload

Other Desires (not requirements)
- Switch from Titan to Ariel for orbital tour inclination changes (*not doable, but increased flybys/optimization is possible*)

*S4. What evidence of exogenic interactions on the surfaces contain? (also S2. composition and M2. plasma sources and dynamics of the magnetosphere)*

**Speakers:** R. Cartwright, C. Grava et al.

**Summary**:

Earth and space-based imaging and spectroscopy of the moons suggest leading/trailing hemisphere differences due to exogenic processing. Because of its likely subsurface ocean, Ariel is a prime focus of study, though all the moons are interesting. An orbiter is required to assess internal oceans, geologic activity, and compositional tracers of ocean-derived materials, though a series of global maps of composition and surface features, as well as higher resolution studies of sites of past/present activity. It is thought that some of the moons may also have seasonally driven exospheres.

Required Measurements:
- Global and spatially-resolved images
    - Regional variations and large-scale geologic features at ~1 km/pixel)
    - Local variations & smaller scale features (~0.1 – 1 km/pixel)
- Reflectance/emission spectra that cover $H_2O$, $CO$, $CO_2$, $CO_3$, $NH_3$/$NH_4$, silicates, organics
- Upstream and local particle populations, including ion composition
- Induced magnetic fields

Lower priority, but enhancing:
- In situ dust sampling proximal to rings and moons
- In situ compositional sampling of exospheres and possible plumes
- Spectral properties and distribution of radiolytic species

Mission Drivers:
- Observations at multiple ranges throughout a flyby
- Global coverage of each major satellite
- Data volume
- Close passes of each moon to search for induced fields (<200 km altitude) and plume environments

Other Desires (not requirements)
- Arial-focused tour (*not doable, but increased flybys/optimization is possible*)
- Arrival before equinox for seasonal change

## Cost-Constrained Mission Exercise

**Description**:

In this exercise, participants were asked to think about what they had heard in the previous sessions and apply it to a mission concept. Everyone was presented with a maximum budget (10$) and set costs for a basic orbiter and probe with minimal payload and tour that cost (6$). The mission included chemical propulsion, with flight time of 12 years, powered by 2 RTGs, requiring instruments to power cycle (no more than 25% powered on time. They could assume Ka-band communication but could only return data every other day. The hardware included a 10-bar atmospheric probe with a mass spectrometer, temp/press sensors only. The orbiter included a 6-filter narrow angle camera, a basic magnetometer, and an IR (point) spectrometer that operated on a 2.5-year orbital tour. The tour included an ~8-month polar orbital phase, rest of the time equatorial with 2 targeted flybys of the 5 major moons only that did not allow radio science and imaging at the same time and with limited close passes of planet or moon.

People were asked to break into multi-disciplinary teams and could add instruments, tour complexity, reduced flight time (with equinox arrival), and more power from a menu with costs to total up to 10$. The ground rules directed them to consider all science disciplines for a broadly-based Flagship. Nothing could be added for free (no contributions could be assumed), and some combinations incurred other penalties/costs. For example, adding too many instruments incurred a power penalty and the need for costing another RTG.

**Outcomes**:

Each group gave a summary of what they chose and why. Most teams wanted to some combination of expanded payload and/or tour length or complexity to enable broader and deeper science. Some added more orbiter instruments, some added additional probe instruments. Nearly all teams wanted a longer orbital tour, including more flybys. Interestingly, no one requested the shorter flight time and equinox arrival, as the additional science to be gained was much more limited than that enabled by the other trades. The exercise was very well-received, and many people more fully appreciated the difficult choices ahead.

## Findings and Next Steps

Overall feedback from the Workshop attendees was very positive. Many people appreciated that the topics were focused and there was a purpose to the meeting beyond presenting the same science as last year. Future workshops should take care to have a focused agenda and not have a meeting with the same people presenting their same usual topics. The collaborative presentation on trajectories from APL, JPL, and GSFC was very well received.

The strongest negative feelings were centered on the poster session. All abstracts were accepted for posters, even those not responsive to the call. This resulted in the poster session being a bit crowded, and presenters felt they did not get as much attention. To help alleviate this, as well as to allow people more time digest the Workshop topics, we will be posting all presentations, written responses to Slido questions not addressed in the sessions, and any submitted posters to an archived website for future reference.

There is obvious concern about getting the mission and studies started soon and how people can participate (with the knowledge that NASA cannot afford to fund a marching army for a decade, or more, before science operations begin). That said, several topics emerged where targeted work can begin sooner or is no longer needed:

- Technology developments:
    - NextGen RTGs (or commitment to several MMRTGs) and the required plutonium availability are a concern.
    - Instrument development for probe instruments under likely entry and coast conditions.
    - Instrument development for cryogenic operation and/or advanced data compression to alleviate power needs on the orbiter.
    - Other technologies that drive mission complexity, risk, power, or cost, such as aerocapture and optical communication, are not required.

- Trajectory work:
    - Optimal cruise trajectories for likely launch windows, launch vehicles, and gravity assists or solar electric propulsion.
    - The orbit insertion can be optimized but depends on the arrival and mission risk assessment (on ring hazards).
    - Tour selection and optimization must wait for payload selection and final science investigation decisions.

- Science prioritization:
    - Once a Project Office is formed, discipline- specific working groups or interdisciplinary scientists can better decide on the threshold science measurements that drive mission choices and tour optimization.

There were several questions about, and interest in, planning a next workshop no sooner than next year. While conveners have yet to come forward, a suggested topic was on instruments – state of art, development needs, and teaming – and this is likely of interest for PESTO to follow. In the meanwhile, we expect science sessions at upcoming conferences.

*Report compiled by A. Simon & K. Mandt based on presentations, workshop discussions, and notes.*